Type of Article: Review

# Title: Imaging mitochondrial calcium dynamics in the central nervous system

Authors: Roman Serrat[1,2,3], Alexandre Oliveira Pinto[1,2],Giovanni Marsicano[1,2,], Sandrine Pouvreau[1,2,]

Affiliations :

[1] Institut National de la Santé et de la Recherche Médicale (INSERM), U1215 NeuroCentre Magendie,

[2] University of Bordeaux, Bordeaux 33077, France.

[3] Present address: Aelis Farma. 33077 Bordeaux, France.

## Abstract:

Mitochondrial calcium handling is a particularly active research area in the neuroscience field, as it plays key roles in the regulation of several functions of the central nervous system, such as synaptic transmission and plasticity, astrocyte calcium signaling, neuronal activity… In the last few decades, a panel of techniques have been developed to measure mitochondrial calcium dynamics, relying mostly on photonic microscopy, and including synthetic sensors, hybrid sensors and genetically encoded calcium sensors. The goal of this review is to endow the reader with a deep knowledge of the historical and latest tools to monitor mitochondrial calcium events in the brain, as well as a comprehensive overview of the current state of the art in brain mitochondrial calcium signaling. We will discuss the main calcium probes used in the field, their mitochondrial targeting strategies, their key properties and major drawbacks. In addition, we will detail the main roles of mitochondrial calcium handling in neuronal tissues through an extended report of the recent studies using mitochondrial targeted calcium sensors in neuronal and astroglial cells, *in vitro and in vivo*.

## Introduction:

Mitochondrial calcium signaling has been the target of a fluctuating research effort since the first hints of its existence back in 1953 (Slater and Cleland, 1953). As described in the historical review of Ernesto Carafoli (Carafoli, 2003), the actual concept of mitochondrial calcium uptake using respiration or ATP-derived energy emerged from the work of two groups in 1961-1962 (Deluca and Engstrom, 1961; Vasington and Murphy, 1962). At that time, indirect measurement of mitochondrial calcium accumulation was achieved by radioactive quantification of the remaining calcium concentration in the supernatant of a mitochondrial suspension. Since then, the subcloning of the main mitochondrial calcium signaling proteins and the development of diverse techniques of evaluation of mitochondrial calcium have sparked a great enthusiasm toward this topic.

Over the years, the main features of mitochondrial calcium handling have been identified, and a plethora of cell functions, including metabolism, cell and inter-organelle communication and cell fate, have been shown to involve mitochondrial calcium. By sequestration and release of calcium, mitochondria play a central role in the shaping of the subcellular dynamics of this universal messenger. For instance, mitochondria strategically positioned at the apical secretory pole of pancreatic acinar cells prevents the propagation of calcium waves to the rest of the cytosol (Tinel et al., 1999). Another example is the tuning of the calcium feedback regulation of calcium channels activity at endoplasmic reticulum-mitochondria contacts (MERCS)(Foskett and Mak, 2010; Williams et al., 2013). Our recent data suggest that this feedback regulation controls the dynamics of astroglial calcium events and, down the line, the integration of synaptic activity by astrocytes (Serrat et al., 2020). Finally, transient increases in mitochondrial calcium levels occur at synapses and regulate neurotransmitter release and synaptic excitability signal (Pivovarova et al., 1999). With such a wide and intricate connection with cell physiology, it is no surprise that a defect in mitochondrial calcium handling is associated with multiple neurodegenerative disorders, such as amyotrophic lateral sclerosis, Alzheimer's disease, Parkinson disease and Huntington disease (Devine and Kittler, 2018; Giorgi et al., 2018). More details on the role of mitochondrial calcium in neuronal and astroglial physiology will be provided at the end of this review.

If multiple approaches do exist to measure mitochondrial calcium content, the most recent developments concern live imaging techniques, and fluorescent probes. One of the main advantages of these tools is that they allow the visualization of mitochondrial calcium movements at high spatial and temporal resolution. It should be underlined, however, that these probes only measure free calcium levels. Mitochondrial calcium can also form phosphate precipitates or be bound to proteins, which will not be detected by dyes but can be revealed by other techniques, such as electron microscopy (see for instance (Wolf et al., 2017) or X ray microanalysis (Pivovarova et al., 1999). With this warning in mind, let's start to explore the world of mitochondrial calcium probes.

## Mitochondrial calcium handling and mitochondrial environment:

Mitochondrial calcium levels are tightly regulated by influx and efflux mechanisms both at the level of the Outer Mitochondrial Membrane (OMM) and the Inner Mitochondrial Membrane (IMM) (Figure 1). Calcium levels in the InterMembrane Space (IMS) are assumed to be similar to the ones in the cytosolic space near mitochondria, while calcium concentrations between 100 nM and 800 µM have been reported in the mitochondrial matrix (David and Barrett, 2003a; Kristián et al., 2002; Montero et al., 2000). In addition, mitochondrial calcium responses can be highly heterogenous, with calcium transients reaching millimolar levels in some mitochondria, probably depending on their spatial proximity with calcium sources (Arnaudeau et al., 2001). Hence, quantification of mitochondrial calcium concentrations using calcium probes can be quite challenging as the sensors will have to respond to a large range of calcium levels. Finally, mitochondria are highly complex and dynamic organelles. The specific mitochondrial environment as well as mitochondrial dynamics can produce imaging artefacts. We will give more details about mitochondrial calcium signaling properties and mitochondrial physiology in the following parts.

### *Mechanisms of mitochondrial calcium influx:*

The 60 years old chemiosmotic theory (Mitchell and Moyle, 1967) provided the thermodynamic basis for rapid accumulation of positively charged ions into the mitochondrial matrix. According to this theory, the mitochondrial electron transport chain (ETC) transfers protons across the low permeability inner mitochondrial membrane (IMM), thereby generating a proton gradient ($\Delta pH_m$) and an electrical gradient ($\Delta\Psi_m$) of -150/-180 mV (negative inside) across this membrane. The $\Delta pH_m$ drives the flow of $H^+$ through the ATP synthase thus the generation of ATP from ADP, and the ionic exchange through electroneutral transporters such as the mitochondrial H+/$Ca^{2+}$ exchanger. Recent studies have revealed that the $\Delta pH_m$ exhibits spontaneous and transient elevations that can be detected by mitochondrial pH sensors, and cause imaging artefacts due to the pH sensitivity of fluorescent proteins (Santo-Domingo and Demaurex, 2012). It is thus important to control $\Delta pH_m$ during imaging experiments to avoid unreliable results. On the basis of the Nernst equation, the $\Delta\Psi_m$ constitutes a large driving force for rapid accumulation of cations, including calcium, in the mitochondrial matrix through passive mechanisms. A proof of this hypothesis was provided by two studies almost 30 years later, that showed that two uncouplers of the oxidative phosphorylation, the dinitrophenol and the carbonyl cyanide-*p*-trifluoromethoxyphenylhydrazone, dissipated the $\Delta\Psi_m$ and abolished mitochondrial calcium uptake (Babcock et al., 1997; Friel and Tsien, 1994).

Before reaching the mitochondrial matrix, calcium crosses the outer mitochondrial membrane (OMM) through the largely unselective Voltage Dependent-Anion Channel (VDAC) (Gincel et al., 2001), and the IMM through another channel, the mitochondrial calcium uniporter (MCU). The VDAC is permeable to solutes smaller than 5kDa. Calcium transfer through VDAC to the mitochondrial IMS depends on cytosolic calcium levels (Báthori et al., 2006). Membrane levels of VDAC might regulate the kinetics of calcium uptake at high calcium concentration microdomains (Rapizzi et al., 2002). However, we will see that calcium entrance into mitochondrial matrix is fast compared to calcium extrusion, due to the nature of the mechanisms involved (O-Uchi et al., 2012).

The MCU is a macromolecular complex which comprises a pore forming subunit and several regulatory proteins (Kamer and Mootha, 2015; Marchi and Pinton, 2014). The pore forming subunit was identified as a 40kDa protein by two different groups in 2011 (Baughman et al., 2011; De Stefani et al., 2011). On the contrary of VDAC, MCUs are highly selective calcium channels (Kirichok et al., 2004). Only a few MCUs are open at physiological calcium levels (100-500 nM), and each channel has a very low conductance (around 0.1 fS), but MCU-mediated calcium uptake increases at higher cytosolic calcium concentration (Boyman and Lederer, 2020). Calcium uptake by the MCU is also regulated by the other proteins of the MCU complex. For instance, MICU3, a protein heavily expressed in the brain, is an enhancer of MCU-mediated calcium uptake (Patron et al., 2019).

Other mechanisms of calcium uptakes through the IMM have been proposed, including IMM-located RyRs, transient receptor potential channel, rapid uptake mode, mitochondrial uncoupling protein 2 and 3, and leucine zipper EF-hand-containing transmembrane protein 1 (LETM1)(Giorgi et al., 2018; O-Uchi et al., 2012). Nevertheless, the MCU is the main mechanism that allows calcium entrance into the mitochondrial matrix.

### *Mechanisms of mitochondrial calcium efflux:*

Following its entrance into mitochondrial matrix, the free calcium either binds to inorganic phosphates (Nicholls, 2005) or is rapidly extruded into the cytosol by a complex system of calcium transporters and channels, restoring the basal state. There are three major systems of extrusion of calcium in the IMM: two antiporters, the mitochondrial $Na^+/Ca^{2+}$ exchanger (mNCX) and the mitochondrial $H^+/Ca^{2+}$ exchanger (mHCX), and one megachannel, the mitochondrial permeability transition pore (mPTP). mNCX appears to be the main antiporter in excitable tissues, including the brain (Giorgi et al., 2018). mNCX function has been credited to NCLX in 2010 (Palty et al., 2010), while the molecular identity of the mHCX remains controversial. The stoichiometry of mNCX-driven calcium transport is electrogenic, with three to four $Na^+$ for one $Ca^{2+}$ (Dash and Beard, 2008; Jung et al., 1995), while the stoichiometry of mHCX is electroneutral (two $H^+$ for one $Ca^{2+}$)(Gunter et al., 1991). mNCX is regulated by calcium, sodium and $\Delta\Psi_{m}$. (Boyman et al., 2013). The maximal mNCX $Ca^{2+}$ flux (18 nmol•mg$^{-1}$•min$^{-1}$) and the $K_m$ values for $[Na^+]_i$ and $[Ca^{2+}]_m$ (~8 mM and 13 µM, respectively) regulate calcium extrusion from mitochondrial matrix, and ultimately shape mitochondrial calcium transient (Boyman et al., 2013).

Under certain conditions, the mPTP can serve as a mitochondrial calcium efflux pathway. Opening of the mPTP causes an abrupt increase of IMM permeability to solutes smaller than 1.5 KDa in mammalian mitochondria (Bernardi and Di Lisa, 2015). The activation of a single of these megachannels can cause the loss of large molecules from the mitochondrial matrix, including synthetic calcium dyes (Wang et al., 2008), and the disruption of mitochondrial osmotic regulation, $\Delta pH_m$ and $\Delta\Psi_M$ (Briston et al., 2019; Carraro et al., 2019; Halestrap, 2010; Wang et al., 2008). All these can cause artefacts while imaging mitochondrial calcium handling.

***Mitochondrial environment and imaging artefacts***

In addition to the pH instability and potential leak of small synthetic probes from mitochondrial matrix, mitochondrial redox signaling and dynamics will add other sources of artefacts that can prevent the accurate detection of mitochondrial calcium movements.

Mitochondria are major sites of redox signaling (Mailloux et al., 2014). They are considered as more reducing than oxidizing compartments, but the presence of numerous exposed thiols makes them highly susceptible to oxidative stress (Kaludercic et al., 2014) hence phototoxicity (Davidson et al., 2007), which can lead to artefactual results. In addition, the mitochondrial electron transfer donors Nicotinamide Adenine Dinucleotide (NAD) and Flavin Adenine Dinucleotide (FAD) can interfere with imaging experiments. FAD fluoresces in the green range while it reduced form FADH(2) is relatively non-fluorescent. On the contrary, NADH fluoresces in the blue range while it oxidized form (NAD) is non fluorescent. Hence, the ratio $\frac{FAD}{NAD(P)H+FAD}$ has been used to measure variations of the mitochondrial redox state (Bartolomé and Abramov, 2015; Kolenc and Quinn, 2019). Both FAD and NADH can be excited with mono photons and multi photons lasers, and display wide excitation and emission spectra. Hence, their emission is difficult to block using barrier filters, and the experimenter should be aware of potential artefacts due to changes in mitochondrial redox status especially with dim and green calcium sensors. Time gated fluorescence signal has been proposed as a way to block more than 95 % of tissue autofluorescence due to FAD and NADH (>20 ns after an excitation pulse) (Chen et al., 2019; Gu et al., 2013; Overdijk et al., 2020). However, this approach is hard to implement with genetically gncoded sensors, as fluorescence proteins display relatively short lifetime (1-4ns) (Billinton and Knight, 2001).

Mitochondria are finally highly dynamics organelles. They move along the axons (Cai and Sheng, 2009) or astrocytes processes (Jackson and Robinson, 2018), which can make them disappear from the field of view. In addition, mitochondria display physiological contraction (Breckwoldt et al., 2014a), which can be mistaken for calcium increase especially if the organelles are smaller than the Z resolution of the imaging system. The shrinkage of the organelle sometimes observed for high calcium concentrations can also affect the proper calibration of the dye (Arnaudeau et al., 2001).

A big part of these artefacts can be addressed by using ratiometric indicators that change the wavelength of the peak of excitation or emission upon binding calcium. This class of dyes permit correction for uneven dye loading, dye leaking, photobleaching and changes in mitochondrial volume. If the use of ratiometric dye is not possible, loading mitochondria with calcium dyes and morphometric dyes will allow the estimation of the signal degradation due to morphological changes of the organelle.

In the following parts, we will see in details the characteristic of the different dyes available to monitor mitochondrial calcium.

## Mitochondrial calcium sensors: synthetic and genetically encoded dyes

A systematic literature search of quantitative measurements of mitochondrial free calcium levels using optical probes retrieves highly different values, sometimes even for the same cells and the same imaging methods (Fernandez-Sanz et al., 2019). One should keep in mind that, in addition to artefacts, high variability can come from the properties of the dyes, the calibration or imaging conditions. Indeed, calcium dyes are not mere light bulbs of certain color that can be fixed inside mitochondria and that turn on and off upon binding/releasing of calcium. They have their own personalities regarding subcellular targeting, affinity to calcium, kinetics of calcium bindings, photostability and sensitivity to mitochondrial environment. Hence, proper choice of the dye and design of the imaging conditions are essential to guarantee the reliability of the results.

Calcium signals are generally short-lived, and free calcium is measured by its binding to the calcium indicator under non-equilibrium conditions. Whether or not the resulting fluorescent transient will faithfully report the “true” calcium transient depends mostly on two characteristics of the dye: its affinity to calcium and its kinetics of calcium binding and unbinding (kon and koff). The affinity of the dye is measured by its Kd, which correspond to the concentration of calcium at which half of the dye molecules are bound to it. The Kd is not a fixed value but varies depending on pH, temperature, medium viscosity, other ions and the binding of the sensor to proteins (Bers et al., 2010). Hence, the value in cells can be quite different from what has been measured in a cuvette, and should be evaluated *in situ* if one wants to measure absolute values of calcium concentration. When possible, sensors should be used to measure calcium concentrations between 0.1 and 10 times their Kd. Indeed, the relationship between the binding of calcium to the indicator and the pCa (logarithm of the free calcium concentrations) follows a sigmoidal shape, centred on the Kd. If the Kd of the probe is too low compared to mitochondrial calcium levels, the probe will be saturated and no further changes in free calcium concentrations will be detected. Changes will also not be detected if the Kd is too high compared to the calcium levels. In addition to the evaluation of the “*in situ* value of Kd”, quantitative measurements of calcium requires a rigorous calibration of the probe. Calibration protocols vary tremendously between laboratories, which will affect the reported values of calcium concentration. If the Kd of the probe controls its capacity to detect changes in calcium concentration, its kinetics alter the recorded calcium transients by filtering the on and off rate of calcium signals.

In addition to the characteristics of the dyes, loading conditions, other fluorophores present in the sample and cell susceptibility to toxicity should be kept in mind while implementing imaging conditions. Indeed, calcium dyes are also calcium buffers that can affect cell physiology at high levels (McMahon and Jackson, 2018), and impact more or less severely the spatio temporal dynamics of calcium signals depending on the mobility, affinity, binding rates and concentration of the sensor (Kovacs et al., 1983; Zhou and Neher, 1993). Calcium probes can also display different cell toxicity and/or phototoxicity. For instance, while comparing mitochondrial calcium signals recorded with different calcium probes, Fonteriz et al. have observed that changes in mitochondrial morphology and membrane potential occur in HeLa cells loaded with synthetic dyes (Rhod-2 and RhodFF) but not with genetically encoded probes.

Synthetic probes (chemically engineered) and genetically encoded sensors are the two main categories of mitochondrial-targeted calcium dyes routinely used for live imaging. In the following parts, we will see the characteristics, advantages and drawbacks of the main representatives of these two categories.

## Mitochondria-targeted synthetic dyes

Synthetic calcium dyes were created in the 1980s by Roger Tsien and collaborators as a modification of the calcium chelator EGTA (Tsien et al., 1984). Over the years, numerous dyes have been engineered, single wavelength-based or ratiometric, with a large range of affinities for calcium (50 nM<Kd > 100 µM) and excitation and emission wavelengths. Most commonly used synthetic calcium probes display binding rate constants between 5 and 20 µs (for review, see (Paredes et al., 2008)). Hence, their kinetics are faster than the ones of any existing genetically encoded calcium probe.

Synthetic calcium dyes are available in three chemical forms: salts, dextran conjugates or acetoxymethyl (AM) esters. Salt and dextran conjugates cannot cross biological membranes (plasma membranes and organelles membranes). AM esters render the calcium dyes hydrophobic, which allow them to cross the cells membranes for non-invasive loading. The AM group is then cleaved by intracellular esterases and the dye trapped inside the cell. A major advantage of using AM-linked calcium dyes is that subcellular compartments can be labelled. For example, low affinity calcium indicators have been used to monitor calcium levels in the endoplasmic reticulum (Launikonis et al., 2006).

### *Targeting of synthetic dyes to the mitochondrial matrix:*

Positively charged AM dyes such as Rhodamine based dyes (Rhod2, X-Rhod1) accumulate preferentially into mitochondria due to their negative $\Delta\psi_m$. Mitochondrial matrix esterases hydrolyse the AM groups, which traps the dye in the mitochondrial matrix .The strategy of the positive charges has also been used to develop other mitochondrial targeted synthetic calcium probes, such as ratiometric ones. In their 2019 article, Diana Pendin and collaborators engineered a fura-2 analogue containing lipophilic triphenylphosphonium (TPP) cations (Pendin et al., 2019). TPP-linked molecules can pass from cell membranes and accumulate into mitochondria (Smith et al., 2003). Complete characterization of the new dye showed minimal incubation time for complete hydrolysis of AM groups of 35 min. Mitochondrial-targeted fura-2 could report calcium transients in HeLa cells, mouse muscle fibres, isolated adult cardiomyocytes and cultured astrocytes.

However, mitochondrial calcium signals detected by synthetic probes are strongly contaminated by the dye remaining in the cytosol. Quenching of Rhod2 by the reducing agent sodium borohydride has been shown to improve the discrimination between cytosolic and mitochondrial signals. The resulting non-fluorescent reduced form of the dye, dihydroRhod2, is oxidized and stains mitochondria (Hajnóczky et al., 1995). However, the dye will spontaneously revert to its oxidized form even in the cytosol of the cell, making the technique not suitable for long recordings. Alternative strategies such as targeted overexpression of esterase (Rehberg et al., 2008) or organelles targeted SNAP-tag fusion proteins (Bannwarth et al., 2009) have been developed for the endoplasmic reticulum and the nucleus, but they have never been tested in mitochondria.

### *Advantages and inconvenient of mitochondrial targeted synthetic dyes:*

The use of synthetic dyes as mitochondrial calcium sensors has many advantages. In addition to the large range of indicators with widely differing Kds that are commercially available, synthetic dyes are linear (Hill coefficient of 1), calibration is easier to achieve than with genetically gncoded dyes (see below), and they are less sensitive to environmental conditions. They also have higher dynamic range and signal to noise ratio than most genetically encoded dyes targeted to mitochondria.

However, synthetic dyes have several drawbacks. Some of them concern all AM dyes, whether they are used to measure cytosolic or organelles calcium levels. Those are incomplete or extracellular AM ester hydrolysis and leakage of the dye from the cell during long recordings. Furthermore, targeting and confinement of the dyes to mitochondria is imperfect. In addition to the cytosolic contamination of the signal. leakage of the dye from mitochondria has been reported for mt-Fura-2 and Rhod2, probably due to opening of the mPTP (Pendin et al., 2019; Wang et al., 2008). Finally, previous studies have reported that members of the Rhod family are toxic for the cells when loaded at concentrations above 2 µm (Fonteriz et al., 2010).

## Mitochondria-targeted genetically encoded calcium Indicators

Genetically encoded calcium indicators (GECIs) are engineered by molecular biology instead of synthetic chemistry. Their older representative, the bioluminescent protein Aequorin, was first introduced in the 1960s for the measurement of intracellular calcium (Ridgway and Ashley, 1967). Fluorescent GECIs became available 30 years after (Miyawaki et al., 1997). Aequorin and other bioluminescent proteins such as obelin contain 3 canonical calcium binding EF hands (Charbonneau et al., 1985; Deng et al., 2005) which allow them to behave as calcium dyes. Fluorescent GECIs were originally created by attaching fluorescent proteins to calcium binding motifs such as calmoduline or troponine (Rose et al., 2014). Over the past 20 years, GECIs have been optimized for calcium affinity, kinetics, stability, and excitation and emission wavelengths (for review, see Looger and Griesbeck, 2012; Nasu et al., 2021; Rose et al., 2014). GECIs can easily be used *in vivo* as they do not require dye loading: the cells themselves synthetize the indicator after being modified by the proper gene. As they are protein-based, GECIs can be targeted to specific cell types by an adequate choice of promoters or to specific cell compartment thanks to the addition of targeting sequences. In the upcoming paragraphs, we will see the different targeting sequences that can be used for targeting GECIs to mitochondria, the bioluminescent and fluorescent probes used in mitochondria, and the advantages and drawbacks of the use of such probes.

### *Mitochondrial targeting sequences*

Most of the work around mitochondrial targeting sequences is related to the mitochondrial matrix. However, mitochondria are complex organelles with multiple compartments that each have specificities in terms of calcium signaling. In the following parts, we will explore the available mitochondrial targeting sequences (illustrated in Figure 2).

*Targeting genetically encoded probes to the mitochondrial matrix*

Two main sequences have been engineered to target first fluorescent proteins, then biosensors, to the mitochondrial matrix. The first one consists in the first 36 amino acids of the subunit VIII of the human cytochrome C oxidase and has been developed by the group of Tullio Pozzan (Rizzuto et al., 1995). We will name it mt8. The second one is from Roger Tsien's group and is composed of the first 12 amino acids of the subunit IV of the yeast cytochrome *c* oxidase or mt4 (Llopis et al., 1998). Unfortunately, it has been observed under several occasions that calcium probes targeted to mitochondria by one replication of these signals only localized efficiently in cells that express very low levels of the construct (Arnaudeau et al., 2001; Filippin et al., 2005; Palmer et al., 2006). In addition, the efficiency of the targeting sequence is probe-dependent. These two points are well illustrated by the study of Filippin et al (2005). The authors reported that, in HeLa cells expressing mitochondrial calcium sensor Camgaroo-2 (targeting sequence mt8) or mitochondrial ratiometric Pericam (targeting sequence mt4), the intensity of the cystosolic fluorescent signal was about 15–20% of the mitochondrial one (Filippin et al., 2005). This was even worse for mitochondrial yellow-Cameleon-2 (targeted with mt8) as the cytosolic fluorescence reached 60% of that of mitochondria on average. Moreover, there was an important variability between cells, with 33% showing only cytosolic signal, while in 5% of the cells the cytosolic fluorescent signal was minor. One should note that if fluorescence intensity is a good marker of the contamination of the detected signal by improperly targeted dye, it is however not a good indicator of the proportion of the probe that is inside mitochondria or the cytosol, as environmental differences as well as a variability in resting calcium levels can affect the intensity of the fluorescence emitted by each molecule of the dye.

To get around the problem of the targeting of sensors to organelles, Miyawaki et al. (Miyawaki et al., 1997) proposed an elegant solution: the “split cameleon”, in which the two functional halves of the calcium sensor are expressed individually with their own targeting sequence. The group of Tullio Pozzan tried to adapt this strategy to mitochondria in 2003 (Filippin et al., 2003), using the sequence mt8. Unfortunately, the authors reported that if the M13-YFP moiety of the sensor was exclusively delivered to the mitochondrial matrix, the targeting of the other half, the CFP-CaM part, was deficient, with ~50% of the fluorescence present throughout the cell body.

The issue was finally solved by the multiplication of the copies of the targeting signal (Filippin et al., 2005; Palmer et al., 2006). In particular, Filippin et al. (Filippin et al., 2005) showed that the use of tandemly duplicated mt4 or mt8 improved the delivery efficacy of all tested genetically encoded probes into the mitochondrial matrix. Palmer et al. (Palmer et al., 2006) later investigated the effect of 2, 4, 6 and 8 repetitions of the targeting sequence mt8 on the localization of their sensors in Hela cells, and demonstrated as well a clear improvement of mitochondrial targeting efficiency by the incorporation of multiple repeats of the tandem signal sequence. Strikingly, the percentage of transfected cells displaying cytoplasmic fluorescence decreased as the number of repeats increased, from 35% for 2mt to 20% for 4mt, 5% for 6mt, and 1% for 8mt.

To finish, it should be noted that the cDNA encoding the signal peptides is fused, in frame, with the cDNA encoding the calcium probes. All calcium probes start by a starting codon embedded in a Kozak sequence. In our hand, mutating the starting codon of the probes (AUG to AGU or methionine to serine) helps with its targeting. We did not further investigate this observation, nor try other conversions.

*Targeting genetically encoded probes to other mitochondrial compartments:*

So far, only a few attempts to target fluorescent indicators to other sub-mitochondrial regions have been made. The first one used the cDNA of the glycerophosphate dehydrogenase (GPD) to address aequorin to the intermembrane space (Pinton et al., 1998). More recently, Giacomello et al (Giacomello et al., 2010) engineered a sensor located at the cytosolic surface of the outer mitochondrial membrane by fusing the first 33 amino acids of the TOM20 N terminal tail (which contain the mitochondrial anchoring signal, (Kanaji et al., 2000) to the calcium probes D1cpV. The resulting sensor revealed mitochondrial calcium hotspots generated by calcium release from the endoplasmic reticulum (ER). To measure calcium signals at MERCS, Csordas et al (Csordás et al., 2010) inserted a RPericam (Nagai et al., 2001) into organelles linkers. The fluorescent probe was targeted by a FK506 binding protein 12 (FKBP12) to the outer mitochondrial membrane, and to the ER by the FKBP-rapamycin binding domain (FRB). Addition of rapamycin formed an OMM-Pericam-FKBP-FRB-ER bridge at the ER-mitochondrial interface, which allow calcium detection at this spot.

More recently, various targeting sequences to the mitochondrial IMS and cristae lumen have been developed by the group of Wolfgang Graier (Waldeck-Weiermair et al., 2019). The authors isolated the first 140 amino acids of the IMS specific protein MICU1 as a targeting sequence to the IMS (the probe is directed to the inner boundary membrane). An inactive mutated version of the recently described protein ROMO1 (reactive oxygen species modulator 1) (Norton et al., 2014) or the subunit e of complex V (CVsue) were used for the cristae lumen. Calcium indicators were the ratiometric GEMGeCO1 (Zhao et al., 2011) and the intensiometric CARGeCO1 (Wu et al., 2013). The three resulting probes exhibited comparable ratio signal under resting conditions, but highly different responses to $IP_3$-mediated calcium mobilization by histamine, with 50% higher increase for the IMS specific probes in comparison to both CL targeted probes, indicating two distinct sub-mitochondrial calcium responses upon calcium release by IP3 receptors. The two calcium indicators within the CL, CVsue-GEMGeCO1 and ROMO1-GEMGeCO1, performed almost identical, but the expression level of the ROMO1 targeted probe was significantly higher enabling a better resolution of the signals. ROMO1 was also used as a targeting sequence for the red shifted sensor GARGeCO1. Hence, one interesting point of this study is that the distinct spectral properties of the green and red GeCO1 biosensors will allow simultaneous calcium recordings in two distinct mitochondrial compartments.

*Side effects of targeting sequences:*

Interestingly, the targeting sequence influences not only the efficiency of the organelle compartmentalization of the probes, but also their functional properties. For instance, Filippin et al (Filippin et al., 2005) have shown that the increase of the ratiometric Pericam (RPericam) 490/415 nm excitation ratio caused by a challenge with histamine in Hela cells was about three-fold larger if the dye was addressed by the 2mt8 sequence compared to the mt4. This could result from an incomplete processing of the targeting sequence, which could affect the fluorescent properties and the calcium affinity of the probe. Indeed, Western Blot analysis demonstrated that the 2mt8-RPericam and the mt4-RPericam are processed differently after their translocation within the mitochondrial matrix, with 2mt8PR being correctly processed while mt4PR remained unmodified and mt8PR was only partially processed (Filippin et al., 2005).

Cell toxicity of the dye is also impacted by the targeting sequence. Indeed, in their 2006 paper, Palmer et al (Palmer et al., 2006) showed that while the probe with 6 or 8 copies of the mt8 had the best localization efficiency, the mitochondria marked with these probe were misshaped (rounded) and a large number of the transfected cells were dying. They thus suggested to use 4mt, as it exhibited a marked improvement in localization without any apparent adverse effects on the mitochondria.

## Bioluminescent probes

Bioluminescence consists in the catalysis of a chemical substrate by a luciferase enzyme, resulting in the production of photons that can be detected by an imaging device (Wang et al., 2018). No excitation light is required for bioluminescence imaging as opposed to fluorescence imaging. Hence, the experimenter will not have to worry about potential phototoxicity, a recurrent problem for mitochondrial imaging as these organelles are very sensitive to light exposure. As mammalian cells lack endogenous bioluminescence, even a small numbers of photons can be detected over background.

The first genetically encoded probe ever used to measure intracellular calcium was the jellyfish Aequora Victoria-derived bioluminescent protein Aequorin (Shimomura et al., 1963). Aequorin is a holoprotein composed of an apoprotein, apoaequorin and a co-factor, coelenterazine (Shimomura et al., 1963). In the presence of calcium ions, colelentarazine is converted to excited coelenteramide and $CO_2$ (Shimomura et al., 1974). Coelenteramide emits blue light (465) while relaxing to the ground state. Hence, one of the major drawbacks of Aequorin is that the probe is consumed by the detection of calcium. Widespread use of Aequorin has also been limited by the complex calcium- and time-dependence of the response, difficulties in the calibration of the light emission in terms of absolute calcium concentration, and the few photons released by the reaction, which imposes very specific detectors and restrains the use of the probe to specific types of calcium signals (Wang et al., 2018).

Nevertheless, Aequorin holds several advantages for mitochondrial calcium imaging, such as a high dynamic range (Kd between 10 and 130 µM with the different mutants), and the assets of the bioluminescent probes which are the lack of phototoxicity and a high signal over noise ratio (Fernandez-Sanz et al., 2019). Aequorin was actually the first mitochondrial-targeted genetically encoded probe ever used to monitor calcium dynamics within the mitochondrial matrix (Rizzuto et al., 1992, 1995), and has since been successfully addressed to the outer mitochondrial membrane or the intermembrane space (Bonora et al., 2013a). Although it first appears as the main weakness of the probe, the fact that it is consumed upon calcium detection can be an asset if experiments are cleverly designed. For instance, in their 2003 study, Filippin et al demonstrated that MERCS are stable over time in HeLa cells. The authors recorded histamine-induced mitochondrial calcium hotspot, reflecting ER calcium release and mitochondrial calcium uptakes at MERCS. The hotspots were systematically smaller during the second histamine stimulation unless the cells were bathed for several minutes in colenterazine, reflecting the consumption of the probe during the first hotspot occurring at the exact same place (Filippin et al., 2003).

If Aequorin is the most common bioluminescent calcium probe to date, other chemiluminescent calcium sensors have been tested, including obelin (Campbell, 1974), photina (Bovolenta et al., 2007), and clytin (Inouye and Tsuji, 1993). Mouse lines have been engineered expressing GFP-Aequorin (Rogers et al., 2007a) and c-Photina (Cainarca et al., 2010a) targeted to mitochondria for *in vivo* imaging. However, few publications were released using these mice. Probably the limitations of these non ratiometric probes, including poor spatial resolution and sensibility to motion artefacts or organelles movements, restrict the use of the mouse lines.

### *Fluorescent probes*

Genetically encoded fluorescent calcium probes can be divided in two categories. The first one, including the Cameleons (Miyawaki et al., 1997, 1999; Romoser et al., 1997; Truong et al., 2001) and the Troponeons (Heim and Griesbeck, 2004), relies on the alteration of the efficiency of fluorescence resonance energy transfer (FRET) between blue or cyan and green or yellow fluorescent proteins upon calcium binding to a calcium responsive linker. The second category uses a single fluorescent protein that contains the calcium sensor as a sequence insert, and comprises the Camgaroos (Baird et al., 1999; Griesbeck et al., 2001), the GCaMPs (Nakai et al., 2001), the Pericams (Nagai et al., 2001), the RCaMPs (Akerboom et al., 2013), the GECOs (Zhao et al., 2011) and the CEPIAs (Suzuki et al., 2014).

The calcium sensing element is usually composed of a calmodulin or one of its mutant version as the calcium binding entity and the calmodulin binding peptide M13. As calmodulin is a ubiquitous signaling protein, the sensor activity could be impaired by post translational modifications of the calmodulin part. In addition, overexpression of calmodulin could affect intracellular signaling. Hence, Heim and Griesbeck designed a novel series of FRET-based indicators, Troponeons, based on troponin, a calcium binding protein found only in skeletal and cardiac muscle (Heim and Griesbeck, 2004). Finally, the group of Jenny Yang tried an alternative strategy, by directly inserting a calcium binding motif EF hand into the GFP in their so-called sensor CatchER (Tang et al., 2011). To our knowledge, mostly calmodulin-based sensors have been targeted to mitochondria.

Genetically encoded fluorescent calcium sensors are plagued with slow kinetics compared to synthetic calcium dyes, albeit to various extents (Rose et al., 2014). Hydrophobic interactions of calmodulin with peptide M13 or slow events close to the troponin EF hands have been identified as the rate-limiting step (Geiger et al., 2012; Rose et al., 2014). Other models take into account the binding of calcium to calmodulin and the fluorescence transfer function to correct the kinetics of the dye (Tay et al., 2007). Calcium binding sites can be mutated to improve the kinetics of the dye, or adjust its Kd to match the calcium levels in the cellular subcompartment of interest. This type of engineering is however challenging, as on- and off rates, affinities and maximal changes in fluorescence intensity are tightly interconnected (Kd = $k_{off}/k_{on}$), and often the optimization of one parameter leads to the degradation of the others (Rose et al., 2014).

Another limitation of genetically encoded calcium sensors results from the use of the calmodulin-M13 hyprid protein, that displays a biphasic calcium binding with two Kds,( 2 µM and 80 nM), resulting from calcium binding to the N- and the C-terminal domains of calmodulin, respectively (Porumb et al., 1994). As a result, certain Cameleons, including Cameleon-1, display a biphasic calcium dependency with apparent dissociation constants of 11 µM and 70 nM, and Hill coefficients of 1.0 and 1.8, respectively (Miyawaki et al., 1997). To suppress this biphasic calcium dependency, the lower Kd has been removed in newer probes by engineering the calcium binding sites (Rose et al., 2014).

*FRET based genetically encoded calcium sensors: The Cameleons*

The first Cameleons targeted to mitochondria were the **Yellow Cameleons**, based on FRET between cyan and yellow proteins (ECFP and EYFP) (Arnaudeau et al., 2001; Miyawaki et al., 1997). The four sensors chosen by Arnaudeau et al. gave a perfect illustration of how the Kd of the dye can affect calcium measurements. Indeed, the authors reported mitochondrial calcium peaks in HeLa cells after histamine application of 3.19 +/- 0.39 µM with YC2 (Kd 1.26 µM), 49.4 +/- 7.4 µM with YC3.1 (Kd 3.98 µM) and 106 +/- 5 µM with YC4.1 (Kd 104 µM) (Arnaudeau et al., 2001).

A more recent Cameleon, **D3cpV**, was then recommended for mitochondrial calcium imaging (Pozzan and Rudolf, 2009a). D3cpV is an improved version of the Yellow Cameleon, in which the calmodulin-peptide pair has been replaced by a mutant version that is unaffected by endogenous calmodulin, and the fluorescent protein YFP by a circularly permuted Venus (cpV), which increase the ratiometric sensitivity and expand the dynamic range of the sensor (Nagai et al., 2004; Palmer et al., 2006). D3cpV has an apparent Kd of 0.6 µM, and a ~5-fold increase in dynamic range. 4mt D3cpV is expressed in the recently developed ROSA26-mt-Cam mouse line which allows mitochondrial calcium measurement *in vivo* (Redolfi et al., 2021). One should be advised however to carefully consider the temporal resolution of this sensor, as our previous work has shown that calcium transients measured with D3cpV display a complex biphasic time course that do not follow a simple two calcium sites binding kinetics (Zhou et al., 2008). Rather, models of the sensor behavior characterize a slow conformational stage that follows calcium binding, with high apparent affinity and unusual [Ca2+] dependence, which suggest a complex reaction that could include a polymerization of the sensor (Zhou et al., 2008). Of note, a newer version of D3cpV has recently been developed for Fluorescence Lifetime IMaging (FLIM) of calcium levels in mitochondria (Greotti et al., 2019). In the new Cameleon, the donor CFP, which displays low fluorescence quantum yield, complex kinetics and a strong tendency to photoswitch, is substituted by the brighter protein mCerulean 3, those single exponential lifetime makes it more suitable for this type of analysis (Greotti et al., 2019).

Red-shifted Cameleons have also been targeted to mitochondria, including the **D1GO-Cam**, which uses green and orange proteins as FRET pair and is compatible with simultaneous recording of cytosolic calcium signals using fura2 (Waldeck-Weiermair et al., 2012).

*Circularly permuted GFP/single fluorescent protein-based calcium sensors*

The engineering of single-fluorescent protein sensors is based on two main discoveries: 1- that large fragments of protein can be inserted into the GFP β barrel without destroying its fluorescence properties (Baird et al., 1999) and 2- that circular permutation, which consist in fusing the N and C termini of a protein with new termini being introduced at another location, is well tolerated when the original N and C termini are fairly close in space, which is the case for GFPs (Heinemann and Hahn, 1995).

The first single fluorescent protein calcium indicators were the **Camgaroos**, in which the Calcium-binding protein calmodulin was inserted between the positions 144 and 146 of a YFP (Baird et al., 1999). Calcium binding to calmodulin and the resulting conformational change caused a seven to eightfold increase in the sensor fluorescence intensity. Apparent Kd for Camgaroo-1 was 7 µM. Variations in the fluorescence intensity of Camgaroos are caused by changes in the protonation state of the fluorescent protein upon calcium binding. Indeed, the protonated species of EYFP, which absorbs at 400 nm, is not fluorescent (Habuchi et al., 2002). The unprotonated form has a peak absorption at 490 nm and is fluorescent. In calcium-free conditions, the construct absorbs predominantly at 400 nm, while in calcium-saturating conditions, the absorption spectrum peaks at 490 nm. Hence, calcium binding shifts the proportion of protonated and unprotonated forms at constant pH, as the pKas for calcium free and calcium saturated dyes are 10.1 and 8.9 respectively. Unfortunately, Camgaroo-1 does not fold well at 37°C and could not be targeted to intracellular organelles, including mitochondria (Baird et al., 1999). The next version, Camgaroo-2, was created by the replacement of Q69K by Q69M in YFP, which lowered its $pK_a$ to 5.7, eliminated its halide sensitivity, doubled its photostability, and improved its folding (Griesbeck et al., 2001). However, mitochondrial-targeted Camgaroo-2 (mtCamgaroo-2) is strongly sensitive to photoconversion when illuminated continuously (with a drop to around 50 % of the initial intensity reported in HeLa cells), and the photoconverted form is insensitive to changes in calcium concentration (Filippin et al., 2003).

The second class of fluorescent protein calcium indicators were the **Pericams**, in which a circularly permuted YFP (cpYFP) has been sandwiched between a calmodulin and a peptide M13 (Nagai et al., 2001). Reversible changes in Pericams fluorescence intensity following fluctuations of ambient calcium concentration also relies on changes in the protonation state of the chromophore upon calcium binding. Several classes of Pericams exist: the Pericams which display an increased fluorescence intensity when they bind to calcium, the inverse

Pericams, those fluorescence decrease calcium binding and the ratiometric Pericams, in which the 400 nm absorption wavelength is fluorescent (Nagai et al., 2001). The ratiometric Pericam has been quickly adopted to measure mitochondrial calcium concentrations (Filippin et al., 2003; Nagai et al., 2001). The use of mitochondrial-targeted ratiometric Pericam (mtRPericam) significantly diminishes the problem of photoconversion that affected mtCamgaroo-2. Filippin et al reported that upon continuous illumination of the mtRPericam, the initial and rapid decrease in fluorescence intensity at both excitatory wavelengths was limited to 15%, and that the initial intensity was recovered after brief periods of non-illumination (Filippin et al., 2003). The significant difference in relation to mtCamgaroo-2 is that the residual fluorescence calcium signal remains sensitive to calcium. It should be noted that *in situ* calibration of mtRPericam retrieved values of 11 µM for Kd as opposed to 1,3 µM *in vitro* (Filippin et al., 2003). One major problem of Pericams, which is common to most insertional mutants of GFP, is their pH sensitivity. In their 2003 study, Filippin et al showed that the 415 nm excitation wavelength of mtRPericam is not as sensitive to pH as the 490 nm one (Filippin et al., 2003). One should note, however, that it is also less sensitive to calcium (Filippin et al., 2003; Nagai et al., 2001). Hence, if mtRPericam constitutes an improvement compared to mt-Camgaroo-2, it should still be used with caution. Of note, mtRPericam and its fluorescent protein, cpYFP, have been used for a decade to report Mitoflashes, increases in cpYFP fluorescence which are not related to calcium fluctuations: they can be recorded at the isoemissive point of the sensor (Pouvreau, 2010; Wang et al., 2008). Mitoflashes have been reported to reflect an increase in superoxide levels and/or an alkalinisation of the mitochondrial matrix (McBride et al., 2019; Pouvreau, 2010; Rosselin et al., 2017; Schwarzländer et al., 2014; Wang et al., 2008, 2016; Wei-LaPierre et al., 2013). Calcium, photoconversion and pH changes are thus probably not the only factors that can change mtRPericam fluorescence intensity.

The third development of single proteins genetically encoded sensors were the **GCaMP**s, which consist in a circularly-permutated GFP inserted between a calmodulin and a peptide M13 (Nakai et al., 2001). The first prototypes suffered from poor expression at mammalian physiological temperature, slow kinetics, but newer versions overcome these problems and display higher baseline fluorescence and large dynamic ranges (Akerboom et al., 2012; Dana et al., 2019; Nakai et al., 2001). GCaMPs have been addressed to mitochondria and successfully used to monitor mitochondrial calcium *in vitro* and *in vivo* (Dana et al., 2019; Lee et al., 2018; Serrat et al., 2020). The calcium-dependence of GCaMP fluorescence intensity is largely driven by an increase in its extinction coefficient, which correspond to a decrease in the pKa of the chromophore upon calcium binding (Akerboom et al., 2009). pH artefacts should thus be considered while monitoring mitochondrial calcium transients. However, as the various versions of GCaMPs display different pKas (Akerboom et al., 2012), one might choose the appropriate one considering the pH of the recorded cell compartment.

GCaMPs formed the basis of more recently developed sensors, such has **RCaMP**, in which **cpGFP has been replaced by the red-shifted cp mRuby (Akerboom et al., 2013), GECO and CEPIA**, engineered by mutation of GCaMPs variants (Wu et al., 2013; Zhao et al., 2011). All three categories of sensors have been successfully used to measure calcium inside mitochondria (Bartok et al., 2019; Suzuki et al., 2014; Wu et al., 2013). GECO and CEPIA come in several color variants, which allow multiplex imaging of different cell compartments or metabolites. If GECO sensors shows a GCaMP-like decrease of pKa from 8.9 to 6.6 upon calcium binding (Zhao et al., 2011), which could lead to artefacts in a mitochondrial environment, RCaMP display a more complex photophysical transition upon calcium binding (Akerboom et al., 2013), and might be a safer alternative.

As new genetically encoded sensors are continuously developed, for instance mNG-GECO1 (Zarowny et al., 2020) or the new GCaMP8 (Zhang et al., 2021), and as the community has becomen aware of the main characteristics that make a sensor a good candidate for mitochondrial targeting, we can predict that other/better sensors will become available in the near future.

*Main drawbacks of genetically encoded fluorescent sensors.*

We have seen in the previous parts that the reliability of the recorded signal can be impaired with genetically encoded fluorescent sensors by artefacts that originate either from the calcium sensing part, or from the fluorescent protein. To summarize, these sensors have Hill coefficient that are lower or higher than 1, resulting in a nonlinear response of the sensor to calcium. In addition, these sensors are pH sensitive, prone to photoswitching, and display complex kinetics. Their behavior is hard to predict *in vitro* as differences in viscosity and concentration of others ions can affect the calcium-binding or the fluorescence properties of the sensor (Yamada and Mikoshiba, 2012). They are also sensitive to heat (Baird et al., 1999). The chromophore formation requires oxygen (Reid and Flynn, 1997), and incomplete chromophore oxidation might lead to an inactive fraction of the biosensor in cells. Finally, signal detection can be impaired by blinking or flickering of the sensor, which can last from nanoseconds to minutes. These fluctuation might be due to triplet states or radical states, proton transfer and conformational dynamics (Yamada and Mikoshiba, 2012).

## Role of mitochondrial calcium in neurons

Mitochondria present a high heterogeneity in the brain, with a vast diversity in cellular mitochondrial content, proteins and functions that depend on the specific cell-type or subcellular localization (Fecher et al., 2019; Lai et al., 1977). These differences correlate with the functional specialization of brain cells, with higher concentration of mitochondria in neurons, which present a high oxidative rate and ion balance homeostasis, than glia (Wong-Riley, 1989). Mitochondrial calcium signaling machinery is also very heterogeneous in brain cells, with observed differences between distinct neuronal or astroglial cell-types in the main mitochondrial calcium channel, the MCU, and the composition of its regulatory subunits (Ashrafi et al., 2020; Fecher et al., 2019). Among the diverse cellular functions of mitochondrial calcium uptake, three are the most-well known representing key features for cell life: (i) modulation of energy production (ATP), (ii) shaping of cytosolic calcium signals and (iii) cell death activation. The modulation of these key cellular roles for neurons and glia will ultimately lead to the control of higher level brain functions such as the the firing rate homeostasis, neurovascular coupling or animal behaviour (Drago and Davis, 2016; Fluegge et al., 2012; Kannurpatti and Biswal, 2008; Ruggiero et al., 2021).

Neurons have limited glycolytic capacity and rely on proper mitochondrial ATP production to maintain ionic gradients and generate axonal and synaptic membrane potentials (Bolaños et al., 2010). Mitochondrial calcium uptake is one of the mechanism that regulates mitochondrial metabolism including the activation mitochondrial dehydrogenases, ATP synthesis and adenylate transporter activity (Denton, 2009; Jouaville et al., 1999; Kennedy et al., 1999; Tarasov et al., 2012; Territo et al., 2000; Wescott et al., 2019). The demonstration of calcium regulation of mitochondrial metabolism was shown among other cells for neurons (Ashrafi et al., 2020; Rangaraju et al., 2014). The precise contribution of mitochondrial calcium to cellular respiration under basal conditions has been debated (Llorente-Folch et al., 2015; Rueda et al., 2014), but recent work has shown that mitochondrial calcium uptake during period of intense neuronal activity is responsible for the triggering of aerobic glycolysis, to ultimately match the increase in energy demand (Díaz-García et al., 2021).

Cytosolic calcium signals are key elements in synaptic transmission and neuronal activity and function. Mitochondria can shape bulk and local cytosolic calcium signals by acting as calcium stores. For example, upon intense neuronal stimulation and consequent calcium entry into the cell, mitochondria can uptake calcium and decrease the amplitude of the bulk cytosolic signal (Billups and Forsythe, 2002; Tang and Zucker, 1997). After neuronal stimulation, when cytosolic calcium levels are decreasing, mitochondria will release the accumulated calcium sustaining long periods of elevated cytosolic residual calcium that will directly influence synaptic transmission and plasticity (Billups and Forsythe, 2002; David and Barrett, 2003b; Lee et al., 2007, 2020; Tang and Zucker, 1997). Further complexity in this regulation arise from the fact that mitochondrial calcium uptake generally occurs at microdomains of high calcium, located around calcium sources, such as from ER (i.e IP3-receptors) and plasma membrane (i.e NMDA receptor) channel. Mitochondrial calcium uptake can potentially regulate the positive and negative feedbacks that the released ion exert on these channels (Foskett and Daniel Mak, 2010; Giacomello et al., 2010; Kannurpatti et al., 2000; Legendre et al., 1993). Of note, mitochondrial calcium uptake was generally considered as a passive mechanism regulated only by calcium concentration (Carafoli, 2003), but recent evidence suggest that this might not be the case in some cell types, including neurons (Ashrafi et al., 2020; Fecher et al., 2019), (Jhun et al., 2016; Marchi et al., 2019; Tarasova et al., 2019). Overall, these different levels of regulation of mitochondrial calcium uptake, together with the discrete distribution of mitochondria along the neurons and their heterogeneity among different neuronal compartments, make the regulation of cytosolic calcium signaling by mitochondria complex and a potential source of diversity which is crucial for the specificity of brain activity.

Mitochondrial calcium signaling is a major spatiotemporal modulator of physiological cytosolic calcium signals. However, mitochondrial calcium overload can be toxic for the cell, ultimately resulting in cell death (Crompton, 1999; Rasola and Bernardi, 2011). This is the case in excitotoxicity, including epilepsy, ischemic conditions or traumatic brain injury, which is induced by the release of high levels of glutamate (Nicholls, 2004). Glutamate, through NMDA and AMPA receptors, is known to over-activate neurons causing and entry of calcium into the mitochondria (Stout et al., 1998). The excessive calcium uptake by mitochondria will disrupt the mitochondrial membrane potential which will ultimately cause a bioenergetics collapse due to blockade of ATP production. Moreover, when concomitant with pro-apoptotic stimuli, mitochondrial calcium overload will produce the opening of the mPTP, a high conductance inner membrane channel, responsible for the release of proapoptotic factors (De Marchi et al., 2014; Rasola and Bernardi, 2011). This, ultimately, will lead to the initiation of the apoptotic cascade and neuronal death (Rasola and Bernardi, 2011).

Due to the importance of mitochondrial calcium dynamics in neuronal physiology and synaptic function, as described in previous points, impaired mitochondrial calcium handling can have dramatic consequences for cell function and survival. Altered calcium signaling, mitochondrial calcium overload and mPTP opening are common features of many neurodegenerative diseases, including Alzheimer disease (Calvo-Rodriguez et al., 2020; Lee et al., 2012), Parkinson's disease (Calì et al., 2012a), Huntington's disease (Choo et al., 2004) or amyotrophic lateral sclerosis (Tradewell et al., 2011). The use of mitochondria-targeted calcium probes has highlighted the temporal correlation of these hallmarks, with mitochondrial calcium overload and mPTP opening being common mechanism preceding cell death (Britti et al., 2018; Calì et al., 2012b).

The first articles analyzing the role of mitochondrial calcium handling in neurons were impacted by the lack of a direct measurement of mitochondrial calcium levels. Nowadays the development of a plethora of calcium sensors specifically targeted to mitochondria allow the experimenter to include direct evaluation of mitochondrial calcium movement to their studies. These sensors must be selected according to the performed experiment and the cellular-type of interest based on the technical criteria required for the particularities of mitochondria such as the Kd, dynamic range or pH sensitivity. In the Table 1 we show a broad sum-up of the main mitochondrial targeted probes and their specific use in different neuronal types. Overall, the new developments in the field of mitochondria-targeted sensors will help to increase our knowledge in the regulation of neuronal calcium signaling and brain function.

## Role of mitochondrial calcium in glial cells

Spatio temporal calcium dynamics are key functional elements of glial cell function. Calcium events in glial cells such as astrocytes and oligodendrocytes are mediated by the activation of inositol 1,4,5-triphosphate receptors (IP3R) causing the release of calcium from the endoplasmic reticulum (ER) (Haak et al., 2001; Sakuragi et al., 2017). The ER is physically located in close apposition to mitochondria (de Brito and Scorrano, 2010; Rizzuto et al., 1998) at contact sites where calcium can transfer from the ER to the mitochondrial matrix. Mitochondrial calcium uptake modulates cytosolic calcium signals and consequently glial function (Boitier et al., 1999; Simpson and Russell, 1996; Smith et al., 2005). Several calcium sensors have been used in glial cells to monitor mitochondrial calcium movement at high spatio-temporal resolution (Table2). Mitochondria are present in glial processes, such as fine astrocytic processes or the myelin sheath where they can control local calcium signaling (Rinholm et al., 2016; Stephen et al., 2015). As in other cell-types, calcium accumulation in glial mitochondria modulates oxidative phosphorylation and energy production (Rizzuto et al., 2012; Wu et al., 2007), although with divergences in the metabolic and calcium uptake properties (Bambrick et al., 2006).

In astrocytes, large populations of mitochondria are found to co-localize with glutamate transporters, dynamically regulating excitatory signaling and brain energetics (Lovatt et al., 2007; Robinson and Jackson, 2016). This regulation can be due in part due to mitochondrial calcium uptake which has been related to vesicular glutamate release (gliotransmission), consequently modulating synaptic communication and excitability (Reyes and Parpura, 2008). Interestingly, glial calcium signaling and function can be also regulated by mitochondrial calcium release. Indeed, it has been demonstrated that calcium release via mitochondrial NCLX is also regulating proliferation and excitotoxic glutamate release in astrocytes (Parnis et al., 2013). Moreover, a role for mitochondria has been suggested in generating calcium microdomains in both astrocytes and oligodendrocytes (Agarwal et al., 2017; Battefeld et al., 2019; Serrat et al., 2020). In addition to the regulation of calcium handling, mitochondrial calcium plays a role in the astrocyte response to injury (Göbel et al., 2020) and mitochondrial calcium overload has been related to apoptotic cell death in both oligodendrocytes and astrocytes (Park et al., 2019; Ruiz et al., 2020).

Less is known about other glial cell types like microglia, where mitochondrial calcium uptake via the mitochondrial transient receptor potential vanilloid 1 channel (TRPV1) depolarizes mitochondria resulting in Reactive Oxygen Species (ROS) production, mitogen activated protein kinase (MAPK) activation, and enhanced migration (Miyake et al. 2015). Neuronal activation was described to increase ATP levels in Schwann cells, causing purinergic receptor activation and mitochondrial calcium increase, which could have an important role for proper myelination (Ino et al., 2015). In summary, the first studies demonstrate a prominent role of mitochondrial calcium in glial cells that will need further exploration to fully understand their contribution to the main glial functions including in brain bioenergetics, myelin formation and synaptic function and remodeling.

## Brain mitochondrial calcium visualization *in vivo*

Most of the experiments performed to observe mitochondrial calcium responses in brain cells have been performed in *in vitro* or *ex-vivo* studies (Table 1). The development of *in vivo* gene delivery techniques such as viral infection, which allows cell-type specific expression of the desired proteins, or the recently development of transgenic mouse, *Drosohpila*, zebrafish or *C.elegans* models carrying mitochondria-targeted sensors, has open the possibility to investigate mitochondrial calcium signaling in living animals (Álvarez-Illera et al., 2020; Esterberg et al., 2014; Garrido-Maraver et al., 2020; Giarmarco et al., 2017; Mandal et al., 2018; Redolfi et al., 2021; Wong et al., 2019). Especially, the visualization of neuronal mitochondrial calcium *in vivo* has been possible thanks to the design of new genetically encoded sensors specifically targeted to mitochondria.

Genetically encoded probes allow the precise labelling of mitochondria and, contrary to the chemical dyes like Rhod2, can be delivered to the cell-type of interest by standard techniques (viral vectors, electroporation or transgenesis) and avoid potential toxic effects due to the hydrolysis of AM groups (Pozzan and Rudolf, 2009b). The first genetically encoded sensors employed for the analysis of mitochondrial calcium were based on bioluminescent probes. However, probes drawbacks, such as a very low spatial resolution, being not appropriate to study single cell mitochondrial calcium dynamics, and the consumption of the co-factor have prevented further development of Bioluminescent-based recordings of mitochondria calcium levels (Bonora et al., 2013b; Ottolini et al., 2014).

Several ratiometric and non-ratiometric mitochondrial targeted fluorescent sensors have been extensively used *in vivo* in *Drosophila* motor nerve terminals or neural stem cells including mitoGCaMP3, mtCamgaroo-2, mtRPericam, 2mt8TN-XXL and Cameleons mtYC2 or 2mt8D4cpV (Chouhan et al., 2010, 2012; Ivannikov and Macleod, 2013; Lee et al., 2018, 2016; Lutas et al., 2012) (Table 1). Zebrafish stably expressing mito-R-GECO in sensory neurons and mito-GCaMP3 in photoreceptor neurons were also created to visualize mitochondrial calcium dynamics *in vivo*. These transgenic fishes allowed to observe in sensory neurons differences between cytosolic/mitochondrial calcium ratio between axons and soma and the impact of mitochondrial retrograde transport inhibition on mitochondrial calcium levels (Mandal et al., 2018, 2021). In photoreceptor neurons, the role of MCU overexpression and knock-out was explored using live larval imaging (Bisbach et al., 2020; Hutto et al., 2020). In mice, the use of mitochondria-targeted calcium fluorescent sensors has been limited to multiphoton imaging after AAV delivery of the sensor. These approaches have been used to visualize the increased mitochondrial calcium levels after Aβ plaque deposition and neuronal death in a transgenic mouse model of cerebral β-amyloidosis expressing mtYC3.6 in the somatosensory cortex (Calvo-Rodriguez et al., 2020), to reveal a CAMKII-regulated probabilistic synchrony between mitochondrial and cytosolic calcium transients in pyramidal neurons from the visual and motor cortex expressing mito-GCaMP6f (Lin et al., 2019).

## Conclusion

In summary, *in vivo* mitochondrial calcium imaging in the brain is still scarce. However, thanks to the development of new live microscopy approaches (such as multiphoton microscopy or fiber photometry), the use of viral delivery techniques or the creation of transgenic animals stably carrying these sensors, we can expect the number of studies dedicated to the visualization of mitochondrial calcium dynamics *in vivo* will increase, including studies of other brain cell types such glial cells or models of neurodegenerative diseases.

## Declaration of competing interest:

The authors declare no competing interest.

## Acknowledgments

This work was funded by: INSERM, European Research Council (Endofood, ERC–2010–StG–260515 and CannaPreg, ERC-2014-PoC-640923, MiCaBra, ERC-2017-AdG-786467), Fondation pour la Recherche Medicale (FRM, DRM20101220445), the Human Frontiers Science Program, Region Nouvelle Aquitaine, Agence Nationale de la Recherche (ANR, NeuroNutriSens ANR-13-BSV4-0006 and ORUPS ANR-16-CE37-0010-01) and BRAIN ANR-10-LABX-0043, to GM; CNRS, Labex Brain (CANNACALC), to SP, and Orups ANR-16-CE37-0010, to R.S.

## Credit author statement:

RS, AOP, GM and SP wrote the paper.

## Tables:

| Probe name | Details | Kd | $\lambda_{ex}/\lambda_{em}$ | Species | References |
|---|---|---|---|---|---|
| Rhod2<br>Rhod5N<br>Rhod-FF | Fluorescent Dye | 0.6 μM/<br>0.32 μM/<br>0.19 μM | 557 nm/<br>581 nm[a] | Mouse/Rat cultured neurons from different brain regions | (Angelova et al., 2019; Britti et al., 2018; Calvo-Rodriguez et al., 2019; Huang et al., 2017; Ludtmann et al., 2019; Peng and Greenamyre, 1998; Qiu et al., 2013; Soriano et al., 2006) |
| | | | | Mouse neuroblastoma N2a cell lines | (Orr et al., 2019) |
| | | | | SH-SY5Y cells | (Paillusson et al., 2017) |
| | | | | Mouse spinal cord motor neurons | (Kruman et al., 1999) |
| | | | | *Droshophila* dopaminergic, larval motorneurons, and brain dissociated neurons | (Ivannikov and Macleod, 2013; Lee et al., 2018, 2016) |
| | | | | Clonal striatal cells | (Quintanilla et al., 2013) |
| | | | | Rat/Mouse spinal cord dorsal horn neurons | (Kim et al., 2011; Takahashi et al., 2019) |
| | | | | Calyx of Held | (Billups and Forsythe, 2002) |
| mtAEQ variants | Bioluminiscence | [b] | 465 nm | Rat cortical or hippocampal cultured neurons | (Baron et al., 2003; Caballero et al., 2016; Calvo-Rodríguez et al., 2016; Sanz-Blasco et al., 2008) |
| | | | | Rat/Mouse cerebellar granule cells | (Caballero et al., 2016; De Mario et al., 2015; Lim et al., 2016; Sanz-Blasco et al., 2008) |
| | | | | Immortalized striatal cells | (Lim et al., 2008, 2016) |
| | | | | *Droshophila* neuroblasts | (Lee et al., 2016) |
| | | | | Mouse sympathetic primary neurons | (Núñez et al., 2007) |
| | | | | SH-SY5Y cells | (Calì et al., 2012a; Hedskog et al., 2013; Zampese et al., 2011) |
| | | | | Mouse neuroblastoma N2a cell lines | (Coussee et al., 2011) |
| | | | | Acute brain slices and *in vivo* bioluminescence brain imaging in newborn mice | (Rogers et al., 2007b) |
| Mito-c-Photina | Bioluminiscence | - | 470 nm | Mouse neural differentiated cells | (Cainarca et al., 2010a, 2010b) |
| | | | | Olfactory sensory neurons from main olfactory epithelium slices | (Fluegge et al., 2012) |
| Mit-GEM-GECO1 | Ratiometric (cpEGFP) | 0.34 μM | 390 nm/ 455 nm-511 nm | Mouse cerebellar neurons | (Lange et al., 2015) |
| | | | | Mouse cortical cultured neurons | (Llorente-Folch et al., 2013; Rueda et al., 2015) |
| Mito-LAR-GECO1.2 | Single FP (cpmApple) | 12 μM | 557 nm/ 584 nm | Mouse dorsal root ganglia neurons | (Wu et al., 2014) |
| Mito-GCaMP2 | Single FP (cpEGFP) | 0.2 μM | 489 nm/ 509 nm | Mouse hippocampal and cortical cultured neurons | (Marland et al., 2016; Qiu et al., 2013) |
| | | 0.542 μM | | *Droshophila* neuromuscular synapse | (Lutas et al., 2012) |
| Mito-GCaMP3 | Single FP (cpEGFP) | | 489 nm/ 509 nm | *Droshophila* neuroblasts | (Lee et al., 2016) |
| | | | | *Drosophila* larval neuromuscular junction, photoreceptor and dopaminergic neurons | (Lee et al., 2018) |
| | | | | Zebrafish photoreceptor neurons in eye live larvae imaging and retinal slices | (Bisbach et al., 2020; Giarmarco et al., 2017; Hutto et al., 2020) |
| Mito-GCaMP5G | Single FP (cpEGFP) | 0.460 µM | 497 nm/ 515 nm | Mouse cortical cultured neurons | (Kwon et al., 2016) |
| Mito-GCaMP6m | Single FP (cpEGFP) | 0.167 µM | 497 nm/ 515 nm | Mouse hippocampal or cortical cultured neurons | (Gazit et al., 2016; Styr et al., 2019; Verma et al., 2017) |
| Mito-GCaMP6f | Single FP (cpEGFP) | 0.375 µM | 497 nm/ 515 nm | Rat/Mouse hippocampal or cortical cultured neurons | (Angelova et al., 2019; Ashrafi et al., 2020; Lin et al., 2019; Naia et al., 2021; Patron et al., 2019) |
| | | | | Excitatory neurons from primary motor cortex or primary visual cortex *in vivo* | (Lin et al., 2019) |
| Mito-Rcamp1h | Single FP (cpmRuby) | 1.6 µM | 570 nm/ 590 nm | Mouse cortical cultured neurons | (Hirabayashi et al., 2017) |
| Mt-Camgaroo-2 | Single FP (cpEYFP variant) | 5.3 µM | 490 nm/ 514 nm | *Drosophila* larval motorneuron terminals | (Chouhan et al., 2010) |
| mtRPericam | Ratiometric (cpEYFP) | 1.7 µM | 415 nm-494 nm/ 517 nm | Rat/Mouse hippocampal or cortical cultured neurons | (Barsukova et al., 2011; Sanmartín et al., 2014; SanMartín et al., 2017) |
| | | | | *Drosophila* larval motorneuron terminals | (Chouhan et al., 2010, 2012) |
| | | | | Rat Preoptic/anterior hypothalamus neurons in slice cultures | (Ikeda et al., 2005) |
| | | | | Rat/Mouse dorsal root ganglia neurons | (Medvedeva et al., 2008; Shutov et al., 2013; Tradewell et al., 2011) |
| Mito-Case12 | Single FP (cpEGFP) | 1 μM | 491 nm/ 516 nm | Mouse hippocampal cultured neurons | (Chang et al., 2011) |
| mtYC2 | FRET (EGFP, EYFP) | 1.24 μM | 433 nm/ 475 nm-527 nm | *Drosophila* larval motorneuron terminals | (Chouhan et al., 2010) |
| mtYC3.1 | FRET (EGFP, EYFP) | 3.98 μM | 433 nm/ 475 nm-527 nm | Rat cortical cultured neurons | (Norman et al., 2007) |
| mtYC3.6 | FRET (EGFP, EYFP) | 0.25 μM | 433 nm/ 475 nm-527 nm | Mouse neuroblastoma N2a cell line and neurons from primary cortical cultures, coronal sections and from somatosensory cortex *in vivo* | (Calvo-Rodriguez et al., 2020) |
| 4mtD1cpv | Ratiometric (ECFP, cpVenus) | - | 433 nm/ 475 nm-528 nm | Mouse/Rat cortical cultured neurons | (Granatiero et al., 2019; Zampese et al., 2011) |
| | | | | SH-SY5Y cells | (Zampese et al., 2011) |
| 4mtD3cpv | Ratiometric (ECFP, cpVenus) | 0.76 μM | 433 nm/ 475 nm-528 nm | Mouse cortical and hippocampal cultured neurons | (Greotti et al., 2019; Qiu et al., 2013; Redolfi et al., 2021) |
| XmtD4cpv[c] | | 49.68 μM | | Rat cortical cultured neurons | (Hill et al., 2014; Ruiz et al., 2018) |
| | Ratiometric (ECFP, cpVenus) | | 433 nm/ 475 nm-528 nm | *Drosophila* larval motorneuron terminals | (Ivannikov and Macleod, 2013) |
| mCerulean3 | Ratiometric (Cerulean3, cpVenus) | 6.2 µM | 433 nm/ 475 nm-528 nm | Neurons from mouse cortical cultures, somatosensotry cortex slices and *in vivo* | (Greotti et al., 2019) |
| mtCEPIA2 | Single FP (cpEGFP) | 0.67 µM | 487nm/ 508nm | Rat hippocampal cultured neurons | (Hotka et al., 2020) |
| Mito-R-Geco | Single FP (cpmApple) | 0.482 µM | 561 nm/ 589 nm | Mouse hippocampal and dorsal root ganglia neurons | (Rysted et al., 2017) |
| | | | | Motor neurons from mouse triangularis sterni explant | (Breckwoldt et al., 2014b) |
| | | | | Mouse neuroblastoma N2a cell line | (Jadiya et al., 2019) |
| | | | | Zebrafish larval posterior lateral line sensory and primary motor neurons | (Mandal et al., 2018, 2021) |
| Mitycam E67Q | Single FP (cpYFP) | 47nm | 498nm/ 515nm | *Drosophila* larval brain neurons | (Garrido-Maraver et al., 2020) |
| | | | | *Drosophila* cultured motor neurons | (Chen et al., 2015) |
| Mt-TN-XXL | FRET (ECFP, cpCitrine) | 60 µM | 433 nm/ 475 nm-529 nm | Mouse cortical cultured neurons | (Huang et al., 2017) |
| | | | | *Drosophila* larval motorneuron terminals | (Ivannikov and Macleod, 2013) |

| | | |
|---|---|---|
| | *Drosophila* larval brain neurons | (Wong et al., 2021) |

**Table 1. Mitochondria calcium sensor used in neuronal cells.** [a] Rhod FF ($\lambda_{ex}/\lambda_{em}$ 553 nm/577 nm) [b] Different affinities according to the variant used [c] "X" denotes for the different number of copies of the mitochondrial target sequence COX VIII. FP: Fluorescent protein.

| Sensor | Cell-type | References |
|---|---|---|
| CEPIA | Astrocytes | (Eraso-Pichot et al., 2017; Larramona-Arcas et al., 2020; Okubo et al., 2019) |
| Mito-R-Geco | Schwann Cells | (Ino et al., 2015) |
| mtAeq | Oligodendrocytes | (Bonora et al., 2014) |
| 2mtD4cpv | Oligodendrocytes | (Ruiz et al., 2020) |
| mtD3cpv | Oligodendrocytes | (Bonora et al., 2014) |
| Rhod2 | Astrocytes | (Boitier et al., 1999; Liu et al., 2009; Ludtmann et al., 2019; Park et al., 2019) |
| Mito-GCaMP6s | Astrocytes | (Li et al., 2014; Montagna et al., 2019, 2019) |
| 4mtD3cpv | Astrocytes | (Redolfi et al., 2021) |
| Rhod5N | Microglia | (Miyake et al., 2015) |
| Rhod5N | Astrocytes | (Britti et al., 2020) |
| Mito-GCaMP5G | Astrocytes | (Li et al., 2014) |
| Mito-GCaMP6f | Astrocytes | (Göbel et al., 2020; Huntington and Srinivasan, 2021) |

**Table 2: Mitochondria calcium sensor used in glial cells**

1 # Figures

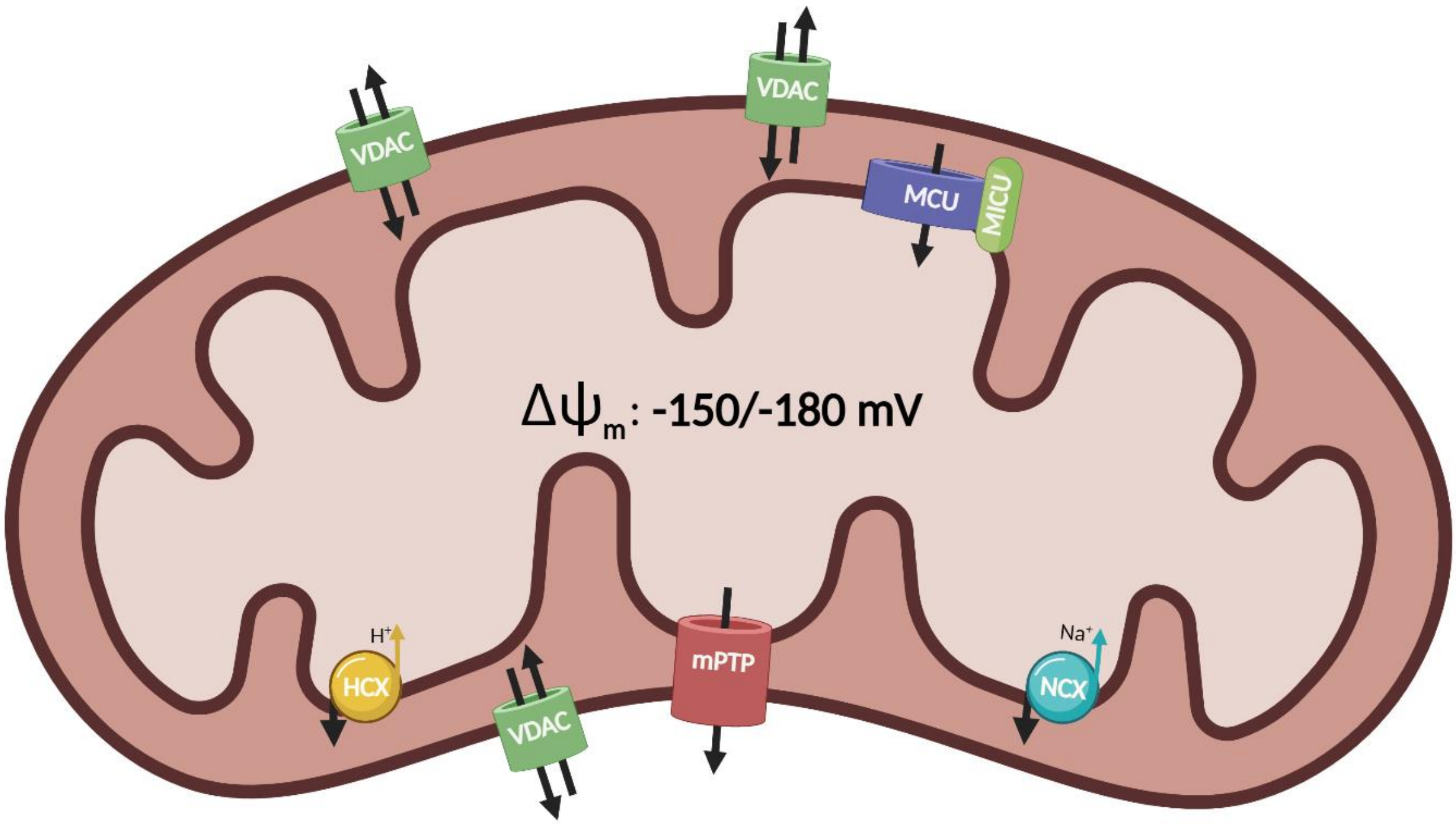

2

3

4 **Figure 1**

5

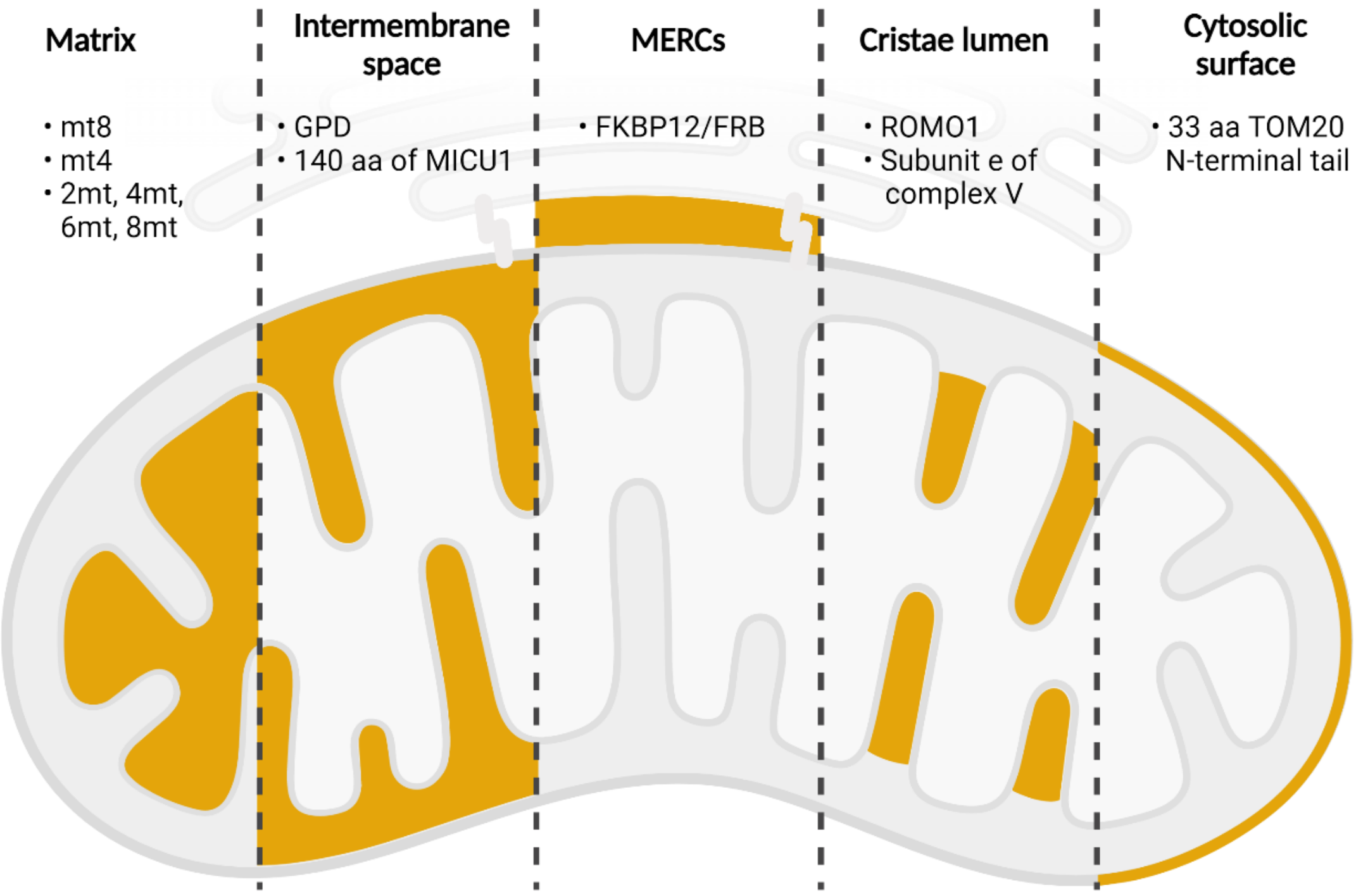
Matrix
• mt8
• mt4
• 2mt, 4mt, 6mt, 8mt
Intermembrane space
• GPD
• 140 aa of MICU1
MERCs
• FKBP12/FRB
Cristae lumen
• ROMO1
• Subunit e of complex V
Cytosolic surface
• 33 aa TOM20 N-terminal tail

**Figure 2:**

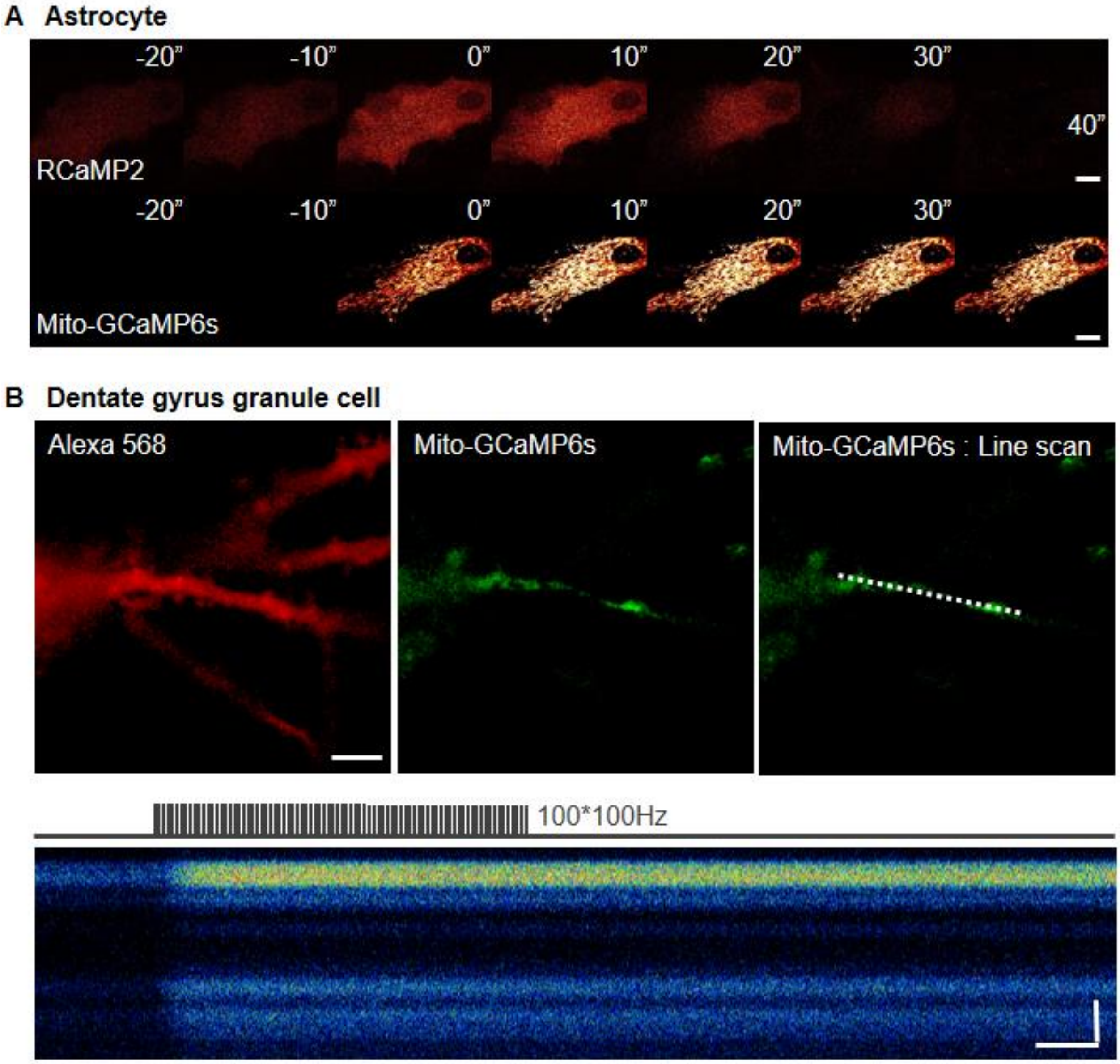

**Figure 3:**

# Legends:

**Figure 1: Mitochondrial calcium handling mechanisms.** Calcium influx is driven by the negative potential of mitochondria ($\Delta\Psi_m$ around -150/-180 mV). **Influx mechanisms**: VDAC: Voltage Dependent Anion Channel, MCU: Mitochondrial Calcium Uniporter, MICU: Mitochondrial Calcium Uptake Protein. **Efflux mechanisms:** HCX: proton/calcium exchanger, NCX: sodium/calcium exchanger, mPTP: mitochondrial Permeability Transition Pore, VDAC: Voltage Dependent Anion Channel.

**Figure 2: Mitochondrial targeting sequences.** aa: amino acids

**Figure 3: Mitochondrial calcium imaging in brain cells: A-Astrocytes:** Confocal images of one astrocyte (primary culture), expressing cytosolic RCaMP2 (upper panels) and Mito-GCaMP6s (bottom panels), at representative time point. ATP treatment (50 µM) was applied at time 0. Scale bar: 10 µm. **B-Dentate gyrus granule cells:** Confocal images of the dendritic tree of a dentate gyrus granule cell (organotypic slice) loaded through a patch pipette with Alexa 568 (right panel) and expressing Mito-GCaMP6s. (middle panel). The left panel shows the line which was recorded during the line scan (bottom panel). A stimulation of 100 short depolarizations at 100 Hz, amplitude 80 mv, duration 2 ms, was applied in voltage clamp mode (holding potential -80 mV). Upper panel: scale bar: 5 µm. Lower panel: vertical scale bar: 5 µm, horizontal scale bar: 0,2 s.